\documentclass[aps,prb,twocolumn,showpacs]{revtex4}
\usepackage{amsmath,amsthm}
\usepackage{epsf}
\usepackage{amssymb}
\usepackage{graphicx}
\usepackage{braket}
\usepackage{etex}
\usepackage{longtable}
\newcommand{\be}{\begin{eqnarray}}
\newcommand{\ee}{\end{eqnarray}}

\newcommand{\nn}{\nonumber\\}

\begin{document}
\title{Random backaction in tunneling of single electrons through nanostructures}
\author{M. Schubotz, T. Brandes}
\affiliation{  Institut f\"ur Theoretische Physik,
  Hardenbergstr. 36,
  TU Berlin,
   D-10623 Berlin,
   Germany}
 \date{\today{ }}
\begin{abstract}
We derive an $n$-resolved  Master equation for quantum transport that includes a dependence on the number $n$ of tunneled electrons in system parameters such as tunnel rates and energy levels. We apply the formalism to describe dynamical changes  due to random backaction effects, for example due to local fluctuations of the electrostatic landscape during the transport process. We  quantify the amount of additional noise on top of electron shot noise due to these fluctuations by giving explicit expressions both for sequential and coherent tunneling examples.
\end{abstract}
\maketitle

 \section{Introduction}

Single electron counting experiments \cite{FCS_exp}
have opened a new view on electronic transport through nanostructures. Fluctuations of the current can provide insight into internal processes that contribute to quantum transport, such as quantum coherent oscillations between different parts of
the nanostructure \cite{Kieetal07}, interactions with the environment, or correlations leading to non-Markovian dynamics \cite{BKF06,Flietal08,MEBA11}.

Counting of individual charges during a stationary transport process  is also fascinating from a fundamental point of view, as it touches the question of measurement
of a quantum system in contact with large reservoirs (source and drain). The standard way to model a counting device in transport is the inclusion of counting fields $\chi$ that appear
in a natural way as Fourier counterparts of $n$, the number of electrons tunneled during a certain period of time $t$.

  Transport Master equations including $\chi$ can be derived in various ways. An early approach stemming from quantum optics is the counting of individual stochastic jump events \cite{Cook81}, e.g.
  in quantum trajectories \cite{Carmichael}. Alternatively, one can start from the projection postulate for $\hat{N}$, a number operator in one of the reservoir, and follow its time-evolution  \cite{Schoenhammer07} which leads to   the Levitov-Lesovik-formula \cite{LL93} for non-interacting, or  $\chi$-dependent Master equations for weakly coupled and interacting systems \cite{EHM09}. Yet another possibility, leading to $n$-resolved Master equations,  is the use of the total system plus reservoir wave function and perturbation theory \cite{GP96}, or the inclusion of additional counter variables in the part of the Hamiltonian that describes the coupling between the system and the reservoirs \cite{SKB09}.

  A somewhat non-trivial aspect  appears when the quantities that determine the transport, such as tunnel rates or energy levels, acquire a dependence on $n$ and thus  become dynamical quantities themselves.
In this paper, we show that such a dependence can in fact be obtained from a  model Hamiltonian that includes appropriate microscopic interaction terms depending on $\hat{N}$, the number operator for one of the reservoirs that acts as a detector. By applying the Nakajima-Zwanzig projection method \cite{Breuer}, we extend the usual derivation of the transport Master equation and obtain an equation of motion for $\rho^{(n)}$, the reduced system density operator conditioned on the charge $n$ in the detector.

We apply our method to model backaction effects that have the form of an  un-controlled  modification of the electrostatic potential landscape
during the transport process. For example, we can account for  saturation effects due to a local piling-up of charge leading to non-constant  tunnel rates $\Gamma_n$ that change with $n$.   Rather than calculating particular model forms of $\Gamma_n$, however, we consider small, but random fluctuations of tunnel barriers and confinement potentials in the course of electrons tunneling through
  the nanostructure and thus assume random, $n$-dependent parameters in the $n$-resolved Master equation.

  Some generic properties of eigenvalue distributions of random transition-rates matrices for rate equations have been studied recently \cite{Timm09}. Our focus here is on
  transport quantities such as the stationary current and the Fano factor in cases of Master equation parameters that randomly change with every tunneled particle. We show that
  the additional noise introduced by this randomness can lead to significant  modifications of current cumulants, in particular in the coherent tunneling regime.

Our formalism is also relevant for the description of
feedback and control operations in the  form of back-action from the detector onto the system during the transport process, whereby the detector is  upgraded into an actuator \cite{KSEB11}.  In feedback control, the Wiseman-Milburn scheme \cite{Wiseman} offers a simple way  to include instantaneous feedback in the form of properly re-defined jump superoperators in the usual transport Master equations.
Other ways to include feedback are possible, however. Our derivations below  provide a microscopic background to the recent combination of open loop control, i.e. a time-dependence in system and reservoir parameters,  with the $n$-dependent output of a counting device to form a feedback loop \cite{Bra10}.

  The outline of this paper is as follows: section II introduces the $n$-resolved transport Master equation via the Nakajima-Zwanzig method. Section III derives the general expression for the ensemble average of the electron current and the Fano factor, and in Section IV we discuss applications to sequential and coherent tunneling through quantum dots.

  \section{Model and Derivation of Master Equation}

  Let us consider a situation where electrons are transfered from a source (`left') reservoir  through a system into a drain (`right')  reservoir.
  We use a decomposition of the total Hamiltonian into system ($S$), bath (left and right reservoir, $B$), and system-bath interaction ($SB$),
  \be
  \hat{{\cal H}}= \hat{{\cal H}}_S +\hat{{\cal H}}_B + \hat{{\cal H}}_{SB}.
  \ee

  The dynamics of the density operator $\varrho$ of the total system (including the reservoirs) is described by the Liouville-von Neumann Equation $\partial_t \varrho= -i[\hat{{\cal H}},\varrho]$.
  We trace out the reservoir degrees of freedom, keeping only the system and the number $n$ of particle transfered from the left into the right reservoir  as relevant degrees of freedom. This can be achieved with the method of Nakajima and Zwanzig \cite{Breuer} by defining a projector $\mathsf{P}$
  \be
  \mathsf{P} \varrho=  \bigoplus_{n}\rho^{(n)} \otimes \varrho_{B,0}^{({n})},
  \ee
  which projects onto the subspace of relevant information. Here,
  \be
  \rho^{(n)} \equiv \operatorname {Tr_{B}^{(n)}} \left({\varrho}\right) \equiv \operatorname{Tr}_{\rm B} (P_n \varrho P_n)
  \ee
  is the reduced system density matrix, conditioned upon the number $n$ of electron transfers.
  Furthermore, the trace operation $\operatorname {Tr_{B}^{(n)}} \left({..}\right)$
  over the bath includes the projection operator $P_n$ that projects onto the subspace of $n$ electrons in the Hilbert space of the right reservoir, and
  \be
  \varrho_{B,0}^{({n})} \equiv \frac{ P_n\varrho_{B,0}P_n }{ \operatorname{Tr}_{\rm B} (P_n \varrho_{B,0}P_n) },
  \ee
  where $\varrho_{B,0}$ is a fixed equilibrium density matrix of the bath. Using the standard Nakajima-Zwanzig technique with the projection operator $\mathsf{P}$, we obtain an equation of motion for the
  projected density matrix to second order in $\hat{\cal H}_{SB}$
  \be\label{NZgeneral}
  & & \partial_t \mathsf{P} \varrho(t) =\mathsf{P} \mathcal{L}_0 \mathsf{P} \varrho(t) \\
& -&\bigoplus_n  \int _0^t dt' \operatorname {Tr_{B}^{(n)}} \left({\left[ \left[ {\mathsf{P} \varrho(t)},{\hat{{\cal H}}_{SB}(t-t')}\right],{\hat{{\cal H}}_{SB}}\right]}\right)\otimes {\varrho}_{B}^{(n)} \nonumber
   \ee
with the Liouvillian $\mathcal{L}_0$ corresponding to system and bath without interactions.

  To make further progress, we need to specify the system-bath interaction $\mathcal{H}_{SB}$. For the following derivation, we assume a simple  Hamiltonian for tunneling between a quantum dot attached to lead reservoirs,
  \be\label{H_tunnel}
  {\cal H}_{SB} = \sum_{k\alpha m}c_{k\alpha}^\dagger {V}_{k\alpha m}(\hat{N})  d_m + H. c.,
  \ee
  where $c_{k\alpha}^\dagger$ is the creation operator for an electron in single particle state $k$ with energy $\varepsilon_{k\alpha}$ in lead $\alpha=$L/R (left/right),
  \be
  {\cal H}_{B}=\sum_{k,\alpha=L,R}\varepsilon_{k\alpha} c_{k\alpha}^{\dagger}c_{k\alpha}^{\phantom{\dagger}}
  \ee
  is the reservoir Hamiltonian,
   and $d_m$ annihilates an electron in system (dot) level $m$. In the tunnel matrix element $V_{k\alpha m}(\hat{N})$ (defined as a Taylor series in $\hat{N}$), we allow a dependence on the number operator $\hat{N}$ of the right reservoir. This dependence describes the `backaction'
  of the right detector reservoir onto the tunnel process and  will render the tunnel rates below to be  $n$-dependent quantities.

  In the following, for simplicity we assume a single system level only and thus suppress the index $m$. The multiple-level case can be done in a completely analogous fashion.    Care has to be taken when evaluating the double commutators in Eq. (\ref{NZgeneral}). We assume the infinite bias limit and thus a bath density matrix
  $\varrho_{B,0}$ corresponding to left and right chemical potentials $\mu_L\gg 0\gg \mu_R$.
  The derivation of the Master equation in the usual Born-Markov approximation is now straightforward algebra, cf. Appendix~A. Setting the single level energy to zero, we arrive at
  \be\label{SRL_master}
   & & \partial_t \rho^{({n})}(t)  =
  \frac{\Gamma_L^{(n)}}{2} \{\rho^{({n})}(t),d d^\dagger \}-\Gamma_L^{(n)} d^\dagger \rho^{({n})}(t) d \nonumber \\
&&+\frac{\Gamma_R^{(n)}}{2} \{      \rho^{({n})}(t),d^\dagger d \}-\Gamma_R^{(n-1)} d  \rho^{({n-1})}(t)  d^\dagger,
  \ee
  where the $n$-dependent tunnel rates are given by the explicit expressions
  \be\label{Gamma_definition}
  \Gamma_\alpha^{(n)}\equiv 2 \pi \sum_k |V_{k\alpha}(n)|^2 \delta(\varepsilon- \varepsilon_{k\alpha}),
  \ee
  and where $\varepsilon_{k\alpha}$ are the single particle energies in the reservoirs $\alpha$. We note that Eq. (\ref{SRL_master}) generalises the usual Master equation for the single resonant level model which is known to be a reliable description of single electron transport \cite{ZSKEB09} in the infinite bias limit employed here.
  The above derivation can now be easily extended to cases where not only the tunnel Hamiltonian ${\cal H}_{SB}$ but also the system Hamiltonian ${\cal H}_S$ has parameters that
  depend on the particle number operator $\hat{N}$ of the counter reservoir.  The $n$-resolved Master equation then has the generic form
  \be\label{nresolvedMaster}
  \partial_t \rho^{({n})}(t)  =\mathcal{L}_0^{(n)} \rho^{({n})}(t) + \mathcal{J}^{(n-1)} \rho^{({n-1})}(t),
  \ee
  where we already used the decomposition of the  Liouvillian into the jump-superoperator $ \mathcal{J}^{(n-1)}$  that describes tunneling of electrons out of the system, and the non-jump part  $\mathcal{L}_0^{(n)}$. Note that the appearance of indices $n$ and $n-1$ reflects the fact that we are within the Markovian approximation (there are no indices $n-2$ etc.) and only consider unidirectional transport (there are no `back' jump terms with index $n+1$).

  \section{Random Ensembles}
  The $n$-resolved Master equation Eq. (\ref{nresolvedMaster}) can in principle be solved easily on a computer once the matrix elements of  the super-operators $\mathcal{L}_0^{(n)}$ and $\mathcal{J}^{(n-1)}$ are known. The easiest situation is, of course, the case where  the super-operators do not depend on $n$ at all, and a simple Fourier transformation of  Eq. (\ref{nresolvedMaster}) according to
  \be\label{rhochieqn}
  \rho_\chi(t) = \sum_n \rho^{({n})}(t) e^{i n \chi}
  \ee
  simply leads to a single, $n$-independent equation for $\rho_\chi(t)$,
  \be
  \partial_t\rho_\chi(t) = \mathcal{L}_0 \rho_\chi(t) + e^{i\chi } \mathcal{J} \rho_\chi(t).
  \ee
  In the following, we assume that with each single electron tunneling event, the complete configuration of the total system changes in a stochastic manner. For example, charges locally piling up near one the tunnel junctions will modify the electrostatic potential of the tunnel barriers and also  affect the confinement potential of the nanostructure.
   In general, these effects will lead to a dynamical modification of tunnel rates or energy levels, depending on $n$, the number of charges tunneled through the structure.
  Usually, this $n$-dependence is small and assumed to be negligeable,  and the electrostatics of the total system is described by means of some average potential landscape thereby giving rise to constant rates and energies.

  A more realistic approach is instead to allow for an $n$-dependent  variation of the parameters in the Master equation. As it is practically impossible to microscopically model the dependence on $n$, the assumption of a {\em random} variation of these parameters should be a good starting point.
  It is therefore physically reasonable that the $n$-dependent super-operators in Eq. (\ref{nresolvedMaster}) are independent,  statistically distributed and uncorrelated to each other. The total Hamiltonian $\mathcal{H}$ underlying  Eq. (\ref{nresolvedMaster}) thereby becomes random, as the Master equation and the density matrix $\rho^{({n})}(t)$ itself. Expectation values
  calculated with the help of $\rho^{({n})}(t)$ then involve {\em two} averages, i.e. the usual trace average (for a given realization within the random ensemble), and the `disorder' (ensemble) average that we denote by $\langle {\dots} \rangle$ in the following.

  The probability $p(n,t)$ of $n$ electrons having tunneled into the right reservoir until time $t$ (counting is started at $t=0$) is given by
  \be
  p(n,t) =\text{Tr} \langle {\rho^{({n})}(t)} \rangle.
  \ee
  As this quantity is in general quite difficult to calculate, we will be satisfied
  with a calculation of mean values and variances in most of what follows.

  \subsection{Ensemble Average}
  To calculate the ensemble average of $\rho^{({n})}(t)$, we first Laplace transform and recursively solve Eq. (\ref{nresolvedMaster}), which leads to
  \begin{align}
  \hat{\rho}_{n}(z) = \hat{\mathcal{P}}^{(n)}(z) \mathcal{J}^{(n-1)} \dots \mathcal{J}^{(1)} \hat{\mathcal{P}}^{(1)}(z) \mathcal{J}^{(0)} \hat{\mathcal{P}}^{(0)}(z)\rho_0,
  \end{align}
  where we defined the propagator
  \be
  \hat{\mathcal{P}}^{(n)}(z)\equiv \left[z-\mathcal{L}_0^{({n})}\right]^{-1},
  \ee
  and $\rho_0$ is an initial  condition. In the long-time limit $t~\to~\infty$ for the stationary state discussed below, the choice of  $\rho_0$ becomes irrelevant in our Markovian theory.

  We now assume the distribution function of the $n$-dependent super-operators in Eq. (\ref{nresolvedMaster}) to factorize into independent and identical distributions for all $n$. This is a reasonable assumption, as long as strong inhomogeneities  are excluded. A counterexample where this assumption fails would be saturation effects, e.g. a gradual and steady
  increase (or decrease) of tunnel rates with increasing number $n$ of tunneled electrons.

  Using the factorization assumption, the {expectation value for the system density matrix in Laplace space} reads
  \begin{eqnarray}
  \langle {\hat{\rho}_{n}(z)}\rangle&=&\hat{\mathcal{W}}(z)^n \hat{\mathcal{P}}(z) \rho_0,
  \end{eqnarray}
  where we defined the ensemble-averaged super-operators
  \be\label{WPdefinition}
  \hat{\mathcal{W}}(z) \equiv \langle \mathcal{J}^{(n)} \hat{\mathcal{P}}^{(n)}(z) \rangle,\quad \hat{\mathcal{P}}(z) =\langle {\hat{\mathcal{P}}^{(n)}(z)}\rangle.
  \ee

Here, the product  $\mathcal{J} \mathcal{P}$ in the definition  of  $\hat{\mathcal{W}}(z)$ is analogous to the product of superoperators  used previously \cite{Bra08} in a discussion of waiting time distributions for Markovian quantum transport.

 Using the Fourier transformation Eq.(\ref{rhochieqn})  with respect to the counting field $\chi$, we obtain the ensemble average
  \begin{align}  \label{eq.rxz.erw}
  \langle {\hat{\rho}_\chi (z)} \rangle= \left[1-e^{i \chi} \hat{\mathcal{W}}(z)\right]^{-1} \hat{\mathcal{P}}(z) \rho_0.
  \end{align}
  From this representation one can now directly obtain the ensemble averaged moments of the Full Counting Statistics in Laplace space, defined as
  \be
  \langle {\hat{m}_{k}(z)}\rangle \equiv \left.  {i^{-k}}\operatorname{Tr} \partial^k_\chi \langle {\hat{\rho}_\chi (z)} \rangle \right|_{\chi \to 0},
  \ee
where we set the electron charge to unity.

  \subsection{Current and Fano Factor}
  The long-time limit for the ensemble averaged electron current  is defined as
  \be
  \langle I_\infty \rangle\equiv\langle {\lim_{t \to \infty} \partial_t m_1(t)} \rangle.
  \ee
  This can be obtained directly in Laplace space via
  \begin{align}
  \langle I_\infty \rangle=\lim_{z \to 0} z^2 \langle \hat{m}_{1}(z) \rangle \label{eq.ILtGen},
  \end{align}
  where here and in the following the hat denotes that the quantity was Laplace transformed, i.e. $\hat f(z) =\int_{0}^{\infty}\mathrm{d}t e^{-zt} {f(t)}$, and the correspondence between  the limit $\lim\limits_{t\to\infty} f(t)$ in time domain and $\lim\limits_{z\to0}z\hat f(z)$ in Laplace space was used.

  Some more effort is needed  to obtain the long-time value of the Fano factor
  \be
  \langle F_\infty \rangle\equiv  \lim_{t\to \infty}  \frac{\langle m_{2}(t)\rangle-\langle m_{1}(t)\rangle ^2}{\langle m_{1}(t)\rangle}.
  \ee
  Here, a direct calculation in Laplace space can not be efficiently executed.
  Fortunately, the Laplace transformation has the convenient property that terms of the order $z^{-k}$ in Laplace space correspond to terms $\frac{t^{n-1}}{n!}$ in the time domain. We thus derive the Fano factor by  first performing an expansion  of $\hat{m}_{2}(z)$ in inverse powers of $z$  up to order $z^0=1$, after which we apply the inverse Laplace transformation in order to  extract the long-time limit.

  This procedure can also be applied in an efficient way for the  higher cumulants \cite{Schubotz11}. In the following, however, we restrict ourselves to the first two cumulants, i.e. the average current and the Fano factor, which we evaluate and discuss for some representative and simple quantum systems.

  \section{Examples}
  We now evaluate our results in detail for single electron tunneling in representative examples of sequential and coherent tunneling.

  \subsection{Tunnel Contact}
  The simplest case is a tunnel contact in which
  the left (source) reservoir is connected to the right (counter) reservoir without anything in between, i.e $\hat{\mathcal{H}}_S=0$, which corresponds to
  super-operators $\mathcal{L}_0^{(n)}=-\Gamma^{({n})} , \mathcal{J}^{(n-1)}=\Gamma^{({n-1})} $ defined by scalar rates $\Gamma^{({n})}$  in our Master equation  Eq. (\ref{nresolvedMaster}).
  The ensemble averages in Eq. (\ref{WPdefinition}) then become
  \be\label{PWdefinition}
   \hat{\mathcal{P}}(z)= \left\langle {\frac{1}{z+\Gamma}}\right\rangle, \quad \hat{\mathcal{W}}(z)=1-z\hat{\mathcal{P}}(z),
  \ee
  where $\Gamma$ denotes the random tunnel rate of the junction.
  Note that $\hat{\mathcal{P}}(z)$ is the generating function for moments of the inverse tunnel rate $\tau \equiv \Gamma^{-1}$,
  \be
  \langle \tau^{k+1} \rangle  =(-1)^{k}\frac{\partial_z^{k} \hat{\mathcal{P}}(z)}{(k)!}.
  \ee
  Using Eq. (\ref{eq.rxz.erw}) and the initial condition $\rho_0=1$ (fixed by the normalization) , we obtain
    \begin{align}
   \langle {\hat{\rho}_\chi (z)} \rangle =\frac{\hat{\mathcal{P}}(z)}{1+e^{i \chi}\left(z \hat{\mathcal{P}}(z) - 1 \right)}.
   \end{align}

  We start with the calculation of the first moment, which evaluates to  $\langle {\hat{m}_{1}(z)} \rangle=\frac{1}{z^2 \hat{\mathcal{P}}(z)}-\frac{1}{z}$, and the long-time limit of the current is therefore
   \begin{align} \label{eq.TAU.GInv}
   \langle I_\infty \rangle=\frac{1}{\hat{\mathcal{P}}(0)}=\frac{1}{\langle {\frac{1}{\Gamma}}\rangle }\equiv{\langle \tau \rangle}^{-1}.
  \end{align}
  This result is interesting in that it yields the current as the inverse of the ensemble averaged time $\tau$ and not, as one might first have expected, the ensemble average
  $\langle \Gamma \rangle$. In the `clean' case, i.e. without random variations and a delta function distribution of the rates, $f(\Gamma') = \delta(\Gamma'-\Gamma)$, the current
  is indeed given by $\langle I_\infty \rangle = \Gamma$, and $\tau=\Gamma^{-1}$ is the  mean value  of the exponential distribution  of waiting times \cite{Bra08} for the Poissonian process describing the tunneling. As we are dealing with single electron tunneling, the waiting time \cite{WEHM08,Bra08,WMY09,AFB11}, rather the tunnel rate,  is in fact the more fundamental quantity to describe the average current:
  if it takes on average a time $\tau $ to send a single object from place A to place B, on average a stationary  current of $1/ \tau $ objects
  flows between A and B. This definition also holds if there are random variations of the time $\tau$ in which case $\tau$ has to be replaced with its ensemble average $\langle \tau \rangle$.

  Next, we calculate the Fano factor, for which we first need the ensemble averaged second moment
  \begin{align}\label{eq.mz2.bar}
   \hat{m}_{2}(z)=\frac{2}{z^3
     \hat{\mathcal{P}}(z)^2}-\frac{3}{z^2
     \hat{\mathcal{P}}(z)}+\frac{1}{z}.
  \end{align}

  We insert the Taylor-expansion of $\hat{\mathcal{P}}(z)$ around $z=0$,  $\hat{\mathcal{P}}(z)\approx \langle \tau \rangle-z\langle \tau^2 \rangle +z^2\langle \tau^3 \rangle+\mathrm O(z^3)$, into   \eqref{eq.mz2.bar} and
  the expression for $\hat{m}_{1}(z)$. After Laplace back-transforming, the moments are then combined to give the Fano factor in the long-time limit,
  \begin{align}\label{Fanosequential}
  \langle F_\infty \rangle= \frac{2 \langle \tau^2 \rangle}{\langle \tau \rangle^2}-1=1+2\frac{\operatorname{Var} (\tau)}{\langle {\tau}\rangle^2},
  \end{align}
  where $\operatorname{Var}(\tau) \equiv \langle \tau^2\rangle -\langle \tau\rangle^2 >0$ is the variance of $\tau$ which thus renders the ensemble averaged Fano factor as greater than unity.

  \subsection{Single Quantum Dot and Ring}
  The tunnel junction example above  can be easily generalized by formally combining $K>1$ junctions in series but still only allowing for one electron tunneling through
  the system at a time. The case $K=2$, for example, corresponds to a single-level quantum dot, a system that has been well studied  experimentally in the past \cite{FCS_exp,Flietal09}.
  In general, a sequence of $K$  tunneling contacts can be regarded as a ring  \cite{Bra08} when the transitions between the first $K$ states, $ 1 \to 2 \dots \to K \to 1$ at rates $\Gamma_i^{(n)}$ are interpreted as no-jump superoperators and the transition $K \to 0$ as the jump process.
  The $n$-dependent Liouvillian super-operators are
  \begin{align}
          \mathcal{L}_0^{(n)}&=\left(
          \begin{array}{ccccc}
           -\Gamma _1^{(n)} & 0 &  0& \dots & 0 \\
           \Gamma _1^{(n)} & -\Gamma _2^{(n)} & 0 &  & 0 \\
           0 & \Gamma _2^{(n)} &   -\Gamma _3^{(n)}  &  &  \vdots \\
           \vdots &  &  & \ddots & 0 \\
            0 & 0 & \dots & \Gamma _{K-1}^{(n)} & -\Gamma _K^{(n)}
          \end{array}
          \right) \nonumber \\
                          \mathcal{J}^{(n-1)}=&\Gamma_K^{(n-1)} \ket{1}\!\rangle\langle\!\bra{K},
  \end{align}
  with $\langle\!\bra{K}=(0,\dots,1)$ and $ \ket{1}\!\rangle=(1,0,\dots)^T $.

  Correspondingly,
  the ensemble averaged super-operators  $\hat{\mathcal{W}}(z)$ and $\hat{\mathcal{P}}(z)$
  now become $K\times K$-matrices,
  \begin{align}
          \hat{\mathcal{W}}(z)&=\left(\begin{array}{*{20}{c}}
          {{W_{\Pi 1}}}&{{W_{\Pi 2}}}&{{W_{\Pi 3}}}& \ldots &{{W_{\Pi K}}}\\
          0&0&0&{}&0\\
          0&0&0&{}& \vdots \\
           \vdots &{}&{}& \ddots &0\\
          0&0& \ldots &0&0
          \end{array}\right)\\
          \hat{\mathcal{P}}(z)&=\left( {\begin{array}{*{20}{c}}
          {{P_1}}&0&0& \ldots &0\\
          {{W_1}{P_2}}&{{P_2}}&0&{}&0\\
          {{W_1}{W_2}{P_3}}&{{W_2}{P_3}}&{{P_3}}&{}& \vdots \\
           \vdots &{}&{}& \ddots &0\\
          {\frac{{{W_{\Pi 1}}{P_K}}}{{{W_K}}}}&{\frac{{{W_{\Pi 2}}{P_K}}}{{{W_K}}}}& \ldots &{{W_{K - 1}}{P_K}}&{{P_K}}
          \end{array}} \right) \nonumber
  \end{align}
  with scalar parameters  $W_i=\langle {\frac{\Gamma_i}{z+\Gamma_i}} \rangle, P_i=\langle {\frac{1}{z+\Gamma_i}}\rangle$ and  $W_{\Pi i}=\prod_{j=i}^K W_j$.
  The disorder averaged moment generating function now has to be defined as usual via the trace operation,
  which for an initially empty system yields
  \begin{align}
  \hat M (i \chi)\equiv\operatorname{Tr} \langle \hat{\rho}_\chi (z)\rangle =	\frac{\sum\limits_{i=1}^K \prod\limits_{k=1}^i P_k}{1-e^{i\chi}W_{\Pi1}}.
  \end{align}
  After back-transformation into the time domain and taking the long-time limit, current and Fano factor read
  \begin{eqnarray}\label{ringcurrentnoise}
  \langle I_\infty \rangle = \frac{1}{\sum\limits_{i=1}^K \langle \tau_i\rangle},\quad
  \langle F_\infty \rangle = \frac{\sum\limits_{i=1}^K(\langle \tau_i\rangle^2+2\operatorname{Var}(\tau_i))}{\left(\sum\limits_{i=1}^K \langle \tau_i\rangle\right)^2}.
  \end{eqnarray}

  Again, the result for the stationary current has its simple origin in an average sequential passage time through the system, for example
  $ \langle I_\infty \rangle=\frac{1}{\langle \tau_L \rangle+\langle \tau_R \rangle}$ for the single level dot ($K=2$). The ensemble averaged Fano factor in this case reads
  \be\label{Fanoaverageseq}
  \langle F_\infty \rangle=\frac{\langle \tau_L\rangle^2 +   \langle \tau_R\rangle^2    }{(\langle \tau_R \rangle+\langle \tau_L \rangle)^2}
  + \frac{2\operatorname{Var} (\tau_L) +2\operatorname{Var} (\tau_R)  }{(\langle \tau_R \rangle+\langle \tau_L \rangle)^2}.
  \ee
  This result has a very natural interpretation: the first term in $\langle F_\infty \rangle$ is identical with the `clean' (non-random) Fano factor for a single level dot with the inverse rates $\Gamma_L^{-1}$ and $\Gamma_R^{-1}$ replaced by the ensemble averaged single-junction waiting times. The second term describes the increase of the current fluctuations that are due to the
  fluctuations of the waiting times $\tau_R$ and $\tau_L$ in the random ensemble.

  We also note that the Fano factor in the sequential tunneling case obeys the inequality
  \be
  \langle F_\infty \rangle \ge  \frac{\sum\limits_{i=1}^K\langle \tau_i\rangle^2}{\left(\sum\limits_{i=1}^K \langle \tau_i\rangle\right)^2} \ge \frac{1}{K},
  \ee
  where the last inequality simply follows from the Cauchy-Schwarz inequality
$( \vec{e}\vec{\tau})^2 \le  \| \vec{e}\| ^2  \| \vec{\tau}\| ^2  = K \| \vec{\tau}\| ^2 $, where $\vec{\tau}$ is the vector with the $K$ components $\langle \tau\rangle_i $ and $\vec{e}$ the vector with the $K$ components $1$.

  \subsection{Double Quantum Dot}
  In contrast to the sequential tunneling examples of the previous section, a coherently coupled system of two quantum dots (double quantum dot, DQD)
  cannot be described as a simple serial combination of single dots \cite{GP96,Gur98,SN96,Bra05}.
  The DQD Hamiltonian
  describes a single additional, spin-polarized  electron on two levels $|L\rangle$ (left dot) and  $|R\rangle$ (right dot)
  that have energy difference $\epsilon$ and are coherently coupled by a matrix element $T_C$.
  Using pseudo-spin Pauli matrices $\hat{\sigma}_z\equiv |L\rangle \langle L| - |R\rangle \langle R|$, $\hat{\sigma}_x\equiv |L\rangle \langle R| + |R\rangle \langle L|$, the Hamiltonian reads
  \be
  {\cal H}&=&{\cal H}_{S}+{{\cal H}_{\rm res}}+{\cal H}_{T}    \\
  {\cal H}_{S}&=&\frac{\epsilon}{2} {\hat{\sigma}_z} + {T_C}(\hat{N})\hat{\sigma}_x ,\quad {\cal H}_{\rm res}=\sum_{k,\alpha=L,R}\varepsilon_{k\alpha} c_{k\alpha}^{\dagger}c_{k\alpha}^{\phantom{\dagger}} \nonumber\\
  {\cal H}_T&=&\sum_{k,\alpha=L,R} (c_{k\alpha}^{\dagger}V_{k\alpha}(\hat{N}) |0 \rangle \langle \alpha|+H.c.)
  \ee
  with the `empty' state $|0\rangle$ and the standard tunnel Hamiltonian ${\cal H}_T$ for coupling to the left and right reservoirs ${\cal H}_{\rm res}$.
  Here, we assume a dependence of the system part $T_C$  on the counting number operator $\hat{N}$ in addition to this dependence in $V_{k\alpha}(\hat{N})$.

  The $n$-dependent superoperators of the DQD in the basis $\rho=(\rho_0,\rho_L,\rho_R,\Re \rho_{RL}, \Im \rho_{RL})$  then have the form
  \begin{align}
          \mathcal{L}_0^{(n)}&=
  \left(
  \begin{array}{ccccc}
   -\Gamma_L^{(n)} & 0 & 0 & 0 & 0 \\
   \Gamma_L^{(n)} & 0 & 0 & 0 & 2 T_{\text C}^{(n)} \\
   0 & 0 & -\Gamma_R^{(n)} & 0 & -2 T_{\text C}^{(n)} \\
   0 & 0 & 0 & -\frac{\Gamma_R^{(n)}}{2} & -\epsilon  \\
   0 & -T_{\text C}^{(n)}  & T_{\text C}^{(n)}  & \epsilon  & -\frac{\Gamma_R^{(n)}}{2}
  \end{array}
  \right), \nonumber \\
  \mathcal{J}^{(n-1)}&=\Gamma_R^{(n-1)} \ket{R}\!\rangle\langle\!\bra{\tilde{R}}
  \end{align}
  with the non-jump superoperator  $\mathcal{L}_0^{(n)}$ and the jump superoperator
  $\mathcal{J}^{(n-1)}$ with $\langle\!\bra{\tilde{R}}=(0,0,1,0,0)$ and $ \ket{R}\!\rangle=(1,0,0,0,0)^T $.
  Furthermore, $T_{\text C}^{(n)}$ is the eigenvalue of $T_C(\hat{N}) $ corresponding to $n$ electrons in the right lead, and $\Gamma_L^{(n)},\Gamma_R^{(n)} $ are $n$-dependent tunnel rates  defined as before  in the sequential tunneling case, Eq.(\ref{Gamma_definition}).

  The  ensemble averaged stationary  current is obtained from the moment generating function (cf. Appendix B, Eq. (\ref{MDQD})) and evaluates to
  \begin{align}\label{IDQD}
  \langle I_\infty \rangle=\frac{1}{\langle \tau_L \rangle+ \langle \tau_R \rangle \left(\epsilon ^2 \langle \tau_T^2 \rangle+2\right)+\tfrac{1}{4}\langle \Gamma_R \rangle \langle \tau_T^2 \rangle},
  \end{align}
  where we defined
  \be\label{DQDaver}
  \langle \tau_L \rangle \equiv \left\langle \frac{1}{\Gamma_L}\right\rangle,\quad
  \langle \tau_R \rangle \equiv \left\langle \frac{1}{\Gamma_R}\right\rangle,\quad
  \langle \tau_T^2 \rangle \equiv \left\langle \frac{1}{T_C^2}\right\rangle.
  \ee
  Some interesting observations can be made from the explicit expressions for $\langle I_\infty \rangle$. First, Eq. (\ref{IDQD}) reduces to the known result \cite{GP96,SN96,EG02}
  in the clean, non-random limit for the parameters $\Gamma_L$, $\Gamma_R$, and $T_C$.

 Second, we recognize that in the expression for $\langle I_\infty \rangle$, the random coupling to the right reservoir enters in the form of two independent averages, i.e. the mean values $\langle\Gamma_R\rangle$ and  $\langle\tau_R\rangle$  of the right tunnel rate and its inverse, respectively. In other words, when fixing the averages  Eq. (\ref{DQDaver}) there is in general no co-incidence between $\langle I_\infty \rangle$ and the corresponding current $I_\infty$ without random flucutations in the parameters. For a random ensemble  of  right tunnel rates $\Gamma_R$,
 the ensemble averaged current $\langle I_\infty \rangle$ is always {\em smaller} than the corresponding clean result $I_\infty $,
  \be\label{alpha_inequal}
  \frac{\langle I_\infty \rangle}{I_\infty } = \frac{1}{1  + \alpha \frac{\langle \tau_T^2\rangle }{4 \langle \tau_R\rangle} I_\infty  } <1,
\quad \alpha \equiv   \left\langle \tau_R\right\rangle  \langle \Gamma_R \rangle {-1}.
  \ee
Here, the  parameter $\alpha$  depends on the probability distribution, but it is always non-negative owing
 to Jensen's inequality     $g(\langle {x}\rangle)\le \langle {g(x)}\rangle$ for convex functions $g(x)$ in the special case $g(x) = \frac{1}{x}$.

\begin{figure}[t]
\includegraphics[width=1\columnwidth]{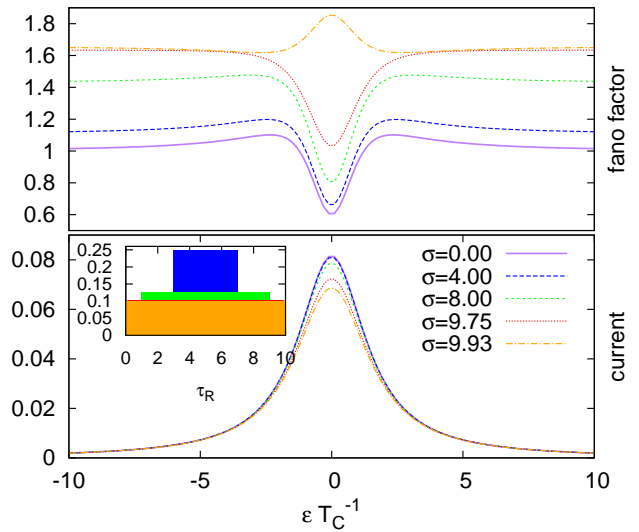}
\caption[]{\label{figure}Stationary currents $\langle I_\infty \rangle$ (lower part, Eq. (\ref{IDQD})), and Fano factors $\langle F_\infty \rangle$ (upper part, Eq. (\ref{Fanodoubleapp})) as a function of internal dot bias $\epsilon$ for non-random case ($\sigma=0$) and
random distributions with width parameter $\sigma>0$ (inset, Eq. (\ref{pdfdouble}))  of inverse tunnel rates $\tau_R=\Gamma_R^{-1}$ at right barrier of double quantum dot. 
Average $\langle \tau_R \rangle = 5$,  other parameters fixed: internal bias $\epsilon$,  tunnel coupling $T_C=\frac{1}{5}$ (in units of left tunnel rate $\Gamma_L$). }
\end{figure}

For example, if we consider  a uniform probability density distribution
\begin{align}\label{pdfdouble}
f_{\sigma,\langle \tau_R\rangle}(x)=\begin{cases}
 \frac{1}{\sigma } & -\frac{\sigma }{2}+\langle \tau_R\rangle\leq x\leq \frac{\sigma
}{2}+\langle \tau_R\rangle \\
 0 & \text{else}
\end{cases}
\end{align}
with average $\langle \tau_R\rangle $ and width $\sigma$ for the inverse right tunnel rate $\tau_R\equiv 1/\Gamma_R$,
we find $ \langle \Gamma_R \rangle  = \frac{1}{\sigma}\log \left(1-\frac{2 \sigma
}{\sigma -2 \langle \tau_R\rangle}\right) .$
For $\sigma\to 2\langle \tau_R\rangle$, i.e.
when arbitrary small right waiting times $\tau_R\to 0$ and thus infinitely large tunnel rates $\Gamma_R$ become
possible, the current $\langle I_\infty \rangle$, Eq. (\ref{IDQD}), becomes more and more suppressed, cf. Fig. \ref{figure}.
This suppression of $\langle I_\infty \rangle$ is a manifestation of the Zeno effect \cite{Bra05}, i.e. strong
detection via the right counter reservoir though at fluctuating detection strengths. It also occurs for other forms of  probability density distributions $f_{\sigma,\langle \tau_R\rangle}(x)$ that give sufficient weight to small  waiting times $\tau_R$.

 Similar to the ensemble averaged current, we also obtain the Fano factor from the moment generating function,
cf. Eq. (\ref{Fanodoubleapp}),
 where second moments with both $\langle \tau_R^2 \rangle$ and $\langle \Gamma_R^2 \rangle$
  enter (in analogy to the parameter $\alpha$, it is possible to introduce positive parameters which describe the
relation between  moments $\langle \tau^{k} \rangle$ and $\langle \Gamma^{k} \rangle$ for $k\ge 1$ \cite{Schubotz11}). 
In contrast to the sequential tunneling result Eq. (\ref{Fanoaverageseq}), one can no longer separate the fluctuations of the current due to the stochastic tunneling process and due to the fluctuations in the tunnel rates, which  is due to the coherent coupling between the two dots.
 Such a separation is only possible at very large internal bias $\epsilon$, where
  \be
  \lim_{\epsilon\to \infty}\langle F_\infty \rangle &=& 1+2  \frac{\text{var} (\tau_R)}{\langle \tau_R \rangle^2} \nonumber \\
  &+&2 \frac{\text{var} (\tau_T^2)}{\langle \tau_T^2 \rangle^2} + 2 \frac{\text{var} (\tau_R)\text{var} (\tau_T^2)}{\langle \tau_R \rangle^2\langle \tau_T^2 \rangle^2}.
  \ee
In this limit, the transport is essentially determined by the single Poissonian process at the right barrier and we recover
  the Fano factor from the sequential tunneling case,  Eq. (\ref{Fanosequential}), plus the additional contribution due to the fluctuations of $T_C^2$ as described by   $\text{var} (\tau_T^2)$.

For smaller $\epsilon$, the Fano factor as a function of $\epsilon$ has no longer a line shape that is simply shifted as compared with the clean (non-random) case, cf. Fig. \ref{figure}. The box distribution example, Eq. (\ref{pdfdouble}), also shows that $\langle F_\infty \rangle$ strongly grows at $\epsilon=0$ when small $\tau_R$ become likely and fluctuations between very long and shorter waiting times become stronger.

\section{Discussion and Conclusion}
For sequential tunneling,  the waiting times $\tau_i$ and their random distributions determine the ensemble averaged transport quantities, which is shown already in the simplest example
of a single tunnel contact where the current is given by the inverse ensemble average $\langle \tau \rangle^{-1}$ and not the averaged rate $\Gamma=\tau^{-1}$, cf. Eq.(\ref{eq.TAU.GInv}).
The random fluctuations of the  $\tau_i$ across the barriers  simply give an additional contribution on top of
the usual shot noise calculated with ensemble averaged waiting times, cf. Eq. (\ref{ringcurrentnoise}).
We expect this correction to be small for all current cumulants, as long as the random fluctuations are small, and experimental results \cite{Flietal09} indeed show that a modeling with fixed Master equation parameters is very successful.

For coherent tunneling, however, the situation seems to be more involved. There can be additional   $n$-dependent fluctuations of internal system parameters (energies, tunnel couplings),
 and we find that fluctuations of the waiting times {\em and} the inverse waiting times (tunnel rates) determine the transport, cf. Eq. (\ref{IDQD}) for the DQD current.
We numerically confirmed that the modifications of DQD current and Fano factor are small if the waiting times $\tau_R$ at the right barrier fluctuate only weakly.
In contrast to the sequential tunneling case, however, the noise due to disorder in the parameters is not simply additive. In particular for
higher cumulants, this might affect a straightforward interpretation of experimental data based on a modeling with Master equations that neglect
the kind of random backaction  discussed here. In contrast, our formalism allows one to take into account these effects, however at the expense
of introducing at least one additional parameter characterizing a distribution function such as the  width $\sigma$ in Eq. (\ref{pdfdouble}). Quantitative statements (such as Eq. (\ref{IDQD}) or Eq. (\ref{Fanodoubleapp})) then require the Liouvillian of the particular nanostructure under consideration.

A further application of our method would be the calculation of noise spectra. An interesting line could be the modeling of $1/f$ charge noise contributings to transport with
dynamically changing tunnel rates $\Gamma_n$.

Finally, another open (though quite challenging) aspect is to go beyond the Markovian scheme used in this paper. One could then test in how far  random parameter fluctuations
effectively would wash out the quantum memory effects in non-Markovian  noise features \cite{MEBA11}.

This work was supported by DFG grant BR 1528/7-1, 1528/8-1, SFB 910, GRK 1558, the Heraeus foundation, and the DAAD. Discussions with W. Belzig are acknowledged.

\begin{appendix}

  \section{$n$-resolved Projection Method for Master Equation}
  The derivation of the Master equation starts from the double commutator expression Eq. (\ref{NZgeneral}) and the form Eq. (\ref{H_tunnel}) for the system-bath interaction $\hat{\cal H}_{SB}$.
  Specifying to a single dot level for simplicity, we obtain the eight terms
  \begin{widetext}
  \begin{eqnarray}
   && {\operatorname{Tr} _B}^{(n)}\left( {\left[ {\left[ {\mathsf{P}\varrho (t),{{\hat H}_{SB}}(t - t')} \right],{{\hat H}_{SB}}} \right]} \right) \\
    &=&   {\operatorname{Tr} _B}^{(n)}\left( {\left[ {\left[ {\mathop  \oplus \limits_{n'} {\rho ^{(n')}} \otimes \varrho _{B,0}^{(n')},\sum\limits_{k\alpha }  \tilde c_{k\alpha }^\dag {{V_{k\alpha }}(\hat N)} {{\tilde d}} + H.c} \right],\sum\limits_{k\alpha } c_{k\alpha }^\dag {{V_{k\alpha }}(\hat N)} {d} + H.c} \right]} \right) \nonumber \\
     &=& {\operatorname{Tr} _B}^{(n)}\left( {\sum\limits_{k\alpha k'\alpha 'n'} {} \varrho _{B,0}^{(n')} \tilde c_{k\alpha }^\dag {V_{k\alpha }}(\hat N){V_{k'\alpha '}^*}(\hat N)} c_{k'\alpha '}^{} \right){\rho ^{(n')}}\tilde dd_{}^\dag  + {\operatorname{Tr} _B}^{(n)}\left( {\sum\limits_{k\alpha k'\alpha 'n'} {} \varrho _{B,0}^{(n')}{V_{k\alpha }^*}(\hat N)\tilde c_{k\alpha }^{}c_{k'\alpha '}^\dag {V_{k'\alpha '}}(\hat N)} \right){\rho ^{(n')}}\tilde d_{}^\dag d \nonumber \\
     \hfill \nonumber \\
     &-& {\operatorname{Tr} _B}^{(n)}\left( {\sum\limits_{k\alpha k'\alpha 'n'} \tilde c_{k\alpha }^\dag {{V_{k\alpha }}(\hat N)}  \varrho _{B,0}^{(n')}{V_{k'\alpha '}^*}(\hat N)} c_{k'\alpha '}^{}\right)\tilde d{\rho ^{(n')}}d_{}^\dag  - {\operatorname{Tr} _B}^{(n)}\left( {\sum\limits_{k\alpha k'\alpha 'n'} {{V_{k\alpha }^*}(\hat N)} \tilde c_{k\alpha }^{}\varrho _{B,0}^{(n')}c_{k'\alpha '}^\dag {V_{k'\alpha '}}(\hat N)} \right)\tilde d_{}^\dag {\rho ^{(n')}}d \nonumber \\
     \hfill \nonumber \\
     &-& {\operatorname{Tr} _B}^{(n)}\left( {\sum\limits_{k\alpha k'\alpha 'n'} {{V_{k'\alpha '}^*}(\hat N)} c_{k'\alpha '}^{}\varrho _{B,0}^{(n')}\tilde c_{k\alpha }^\dag {V_{k\alpha }}(\hat N)} \right)d_{}^\dag {\rho ^{(n')}}\tilde d - {\operatorname{Tr} _B}^{(n)}\left( {\sum\limits_{k\alpha k'\alpha 'n'} c_{k'\alpha '}^\dag {{V_{k'\alpha '}}(\hat N)}  \varrho _{B,0}^{(n')} {V_{k\alpha }^*}(\hat N)} \tilde c_{k\alpha }^{} \right)d{\rho ^{(n')}}\tilde d_{}^\dag  \nonumber \\
     \hfill \nonumber \\
     &+& {\operatorname{Tr} _B}^{(n)}\left( {\sum\limits_{k\alpha k'\alpha 'n'} {{V_{k'\alpha '}^*}(\hat N)} c_{k'\alpha '}^{}\tilde c_{k\alpha }^\dag {V_{k\alpha }}(\hat N)\varrho _{B,0}^{(n')}} \right)d_{}^\dag \tilde d{\rho ^{(n')}} + {\operatorname{Tr} _B}^{(n)}\left( {\sum\limits_{k\alpha k'\alpha 'n'}  c_{k'\alpha '}^\dag {{V_{k'\alpha '}}(\hat N)}  {V_{k\alpha }^*}(\hat N) \tilde c_{k\alpha }^{}\varrho _{B,0}^{(n')}} \right)d\tilde d_{}^\dag {\rho ^{(n')}},  \nonumber
  \end{eqnarray}
  \end{widetext}
  where the tilde in the operators abbreviates the time-dependence, e.g. $\tilde c_{k\alpha } = c_{k\alpha } e^{-i\epsilon_{k\alpha}(t-t')} $ where
  $\epsilon_{k\alpha}$ is a single-particle energy in lead $\alpha$, and $\rho ^{(n')}$ depends on $t'$.
  We  assume the infinite bias limit and thus a bath density matrix
  $\varrho_{B,0}$ corresponding to left and right chemical potentials $\mu_L\gg 0\gg \mu_R$.
  The bath correlation function for the jump term for the right (counter)  reservoir, e.g., is then determined by
  \begin{eqnarray}\label{Trace_example}
  & &\operatorname{Tr}_{\rm B}( P_n    c_{k'\alpha'}^\dagger  V_{k'\alpha'}(\hat{N})  \varrho^{(n')}_{B,0} {V}_{k\alpha}^* (\hat{N}) c_{k\alpha}   P_n ) \nonumber\\
  &=&\delta_{n'n-1}\delta_{kk'}|V_{k\alpha}(n-1)|^2  \delta_{\alpha,R},
  \end{eqnarray}
  where we kept in mind that $P_n$ projects onto the subspace of $n$ electrons in the  right (and not the left) reservoir $\alpha=R$, and $\hat{N}$ in $V_{k'\alpha'}(\hat{N})$ is the number operator of the right reservoir. The expression Eq. (\ref{Trace_example}) is still approximate in the sense that it neglects the difference between the usual grand-canonical Fermi functions $f_{kR} = \operatorname{Tr}_{\rm B} \varrho_{B,0} c_{kR}^\dagger
c_{kR}$ and the canonical projections of the occupations at fixed particle number $n$. This difference, however, becomes irrelevant for $\mu_R\to -\infty$, or more generally, if for arbitrary  chemical potential $\mu_R$ the projection $P_n$ is on a subspace with $n+N_0$ particles, where $N_0$ is a macroscopic particle number determined by $\mu_R$, in which case
the $\alpha=R$ term in Eq. (\ref{Trace_example})  would have to be replaced by the Pauli block factor $1- f_{kR}$.

  A non-jump-term associated with  the left (source) reservoir (again for $\mu_L\gg 0\gg \mu_R$) has the form
  \begin{eqnarray}
  & &\operatorname{Tr}_{\rm B}( P_n    \varrho^{(n')}_{B,0}c_{k'\alpha'}^\dagger V_{k'\alpha'}(\hat{N}) {V}_{k\alpha}^* (\hat{N})  c_{k\alpha}  P_n)\nonumber\\
  &=&\delta_{nn'}\delta_{kk'}|V_{k\alpha}(n)|^2  \delta_{\alpha,L},
  \end{eqnarray}
  where we used the fact that $\hat{N}$ commutes with $c_{kL}^{(\dagger)}$. Crucially, the tunneling matrix elements now depend on the eigenvalues $n$ of the number operator $\hat{N}$.
  In the infinite bias limit, the Markov approximation becomes exact in the integration over time $t'$, and the time dependence in the Fermion operators leads to a delta function that together with the sum over $k$ yields the $n$-dependent tunnel rates $\Gamma_\alpha^{(n)}$, Eq. (\ref{Gamma_definition}).

\section{Moment Generation Function and Fano Factor for Double Dot}
  Using the techniques outlined in section III, we perform a random ensemble average where we again assume uncorrelated distributions for  $\Gamma_L^{(n)}$, $\Gamma_R^{(n)} $ and $T_{\text C}^{(n)}$ that are independent for all $n$.  The calculation of the moment generation function is analogous to the calculation for the sequential case, with the result
  \be\label{MDQD}
  M(i\chi)&=&\textrm{Tr}\left(\langle[z-\mathcal{L}_{0}^{(n)}]^{-1}\rangle\frac1{1-e^{i\chi}\langle\frac{\mathcal{J}^{(n)}}{z-\mathcal{L}_{0}^{(n)}}\rangle}\rho_{0}\right) \nonumber\\
  &=& \frac{\langle {A(z)}\rangle -W_L(z) \langle{ B(z)}\rangle -\frac{1}{4}P_L(z) }{e^{i \chi} W_L(z) \langle{\Gamma_R B(z)}\rangle},
  \ee
  where again an initially empty dot was assumed and
  \be
   P_L(z)\equiv \langle {\frac{1}{z+\Gamma_L}} \rangle, \quad W_L\equiv 1-zP_L(z)
  \ee
  refer to the left tunnel barrier and are defined in analogy to Eq. (\ref{PWdefinition}), whereas
  \begin{align}
  A(z) =&\frac{(\Gamma_R+2 z) \left((\Gamma_R+z) (\Gamma_R+2 z)+4 T_{\text C}^2\right)+4 \epsilon ^2 (\Gamma_R+z)}{(\Gamma_R+2 z)^2 \left(z (\Gamma_R+z)+4 T_{\text C}^2\right)+4 z \epsilon ^2 (\Gamma_R+z)},\nn
  B(z) =& T_{\text C}^2 \frac{\Gamma_R+2 z}{(\Gamma_R+2 z)^2 \left(z (\Gamma_R+z)+4 T_{\text C}^2\right)+4 z \epsilon ^2 (\Gamma_R+z)}
  \end{align}
  are combinations containing the random variables $\Gamma_R$ and $T_C$ that have to be averaged over in the definition  Eq. (\ref{MDQD}).
Using these result, we obtain the stationary, ensemble averaged  Fano factor for the double quantum dot
\begin{widetext}
  \be\label{Fanodoubleapp}
  {\langle F_\infty \rangle}&=&\frac{1}{\langle I_\infty \rangle^2}\Big[-\langle \tau_L \rangle^2+2 \langle \tau_L^2 \rangle +\langle \tau_T^2 \rangle \left(-4 \epsilon ^2 \langle \tau_R \rangle^2+10 \epsilon ^2 \langle \tau_R^2 \rangle-\langle \Gamma_R \rangle \langle \tau_R \rangle+\frac{1}{2}\right)\\
  &-&\langle \tau_T^2 \rangle^2 \left(\epsilon ^4 \langle \tau_R \rangle^2+\frac{1}{2} \epsilon ^2 \langle \Gamma_R \rangle \langle \tau_R \rangle+\frac{1}{16} \langle \Gamma_R \rangle^2\right)+\langle \tau_T^4 \rangle \left(2 \epsilon ^4 \langle \tau_R^2 \rangle+\frac{1}{8} \langle \Gamma_R^2 \rangle+\epsilon ^2\right)-4 \langle \tau_R \rangle^2+8 \langle \tau_R^2 \rangle\Big]\nonumber.
  \ee
  \end{widetext}

\end{appendix}


 \end{document}